\documentclass[12pt]{revtex4}
\usepackage{graphicx}
\usepackage{wrapfig}
\usepackage{dcolumn}
\usepackage{bm}
\makeatletter
\def\@dotsep{4.5}
\makeatother

\begin{document}
\newcommand{\bR}{{\bf R}}
\newcommand{\bbr}{{\bf r}}

\newcommand{\bP}{{\bf P}}

\title{ Transition metal oxides using quantum Monte Carlo}
\author{Lucas K. Wagner \footnote{Present address: 
366 Le Conte Hall \#3700; Berkeley, CA 94720; lkwagner@berkeley.edu}}

\address{Center for High Performance Simulation and Department of Physics \\ 
North Carolina State University, Raleigh, NC 27695}

\date{\today}

\begin{abstract}
The transition metal-oxygen bond appears prominently throughout chemistry and solid-state physics.  Many materials, 
from biomolecules to ferroelectrics to the components of supernova remnants contain this bond in some form.  Many of these materials' properties strongly depend on fine details of the TM-O bond and intricate
correlation effects, which make accurate calculations of their properties very challenging.
  We present quantum Monte Carlo, an explicitly correlated
class of methods, to improve the accuracy of electronic structure calculations 
over more traditional methods like density functional theory.  We find that unlike s-p type bonding,
the amount of hybridization of the d-p bond in TM-O materials is strongly dependant on electronic correlation.
\end{abstract}
\maketitle

\section{Introduction}

Transition metal chemistry is a particularly exciting area of research, with
 applications from astrophysics to biology
to potential inexpensive high-efficiency solar cells and high-temperature 
superconductivity.  Because of the partially 
filled d-shell, transition metals can 
form many types of bonds and can also exhibit ferroelectric and ferromagnetic 
ordering.  Transition metal oxides
are particularly interesting because they are one of the most common 
transition metal complexes, and  exhibit most of the above effects.  This rich physics is
quite difficult to describe theoretically, however, since electronic correlation is very strong in these
materials.  Current approximate density functional theories tend
 to perform quite poorly 
on transition metals, particularly in comparison to its quite good accuracy on 
elements with s and p type bonding.  Problematic quantities are not hard to find; they 
include the dipole moment in molecules, binding (or cohesive energies), the 
lattice constants of perovskites, high pressure behavior, and band gaps/excitation energies.

Rather than attempting to improve the approximate density functional, quantum Monte
 Carlo (QMC) approaches  
take a different direction--explicitly treating the electronic correlation in a wave
 function based approach, while maintaining reasonable scaling with system size.
It can be made to scale from O(1) to O(N$^3$) in the number of electrons\cite{qmc_ordern}, depending
on the quantity of interest.
QMC attains very low upper-bound energies on medium-sized electronic problems (up to 
thousands of electrons at the time of writing), and has been used as a benchmark method
on s-p systems\cite{jeff_benchmark}.  Since it 
treats the electronic correlation explicitly in the many-body wave function, it is a
promising method for strongly correlated TMO systems.

The goal of this review is to summarize the current state of the art of QMC as 
applied to TMO's.  This is a fairly new field, with few calculations.  Most of 
these calculations have benchmarked the method to determine
the accuracy that one should expect.  This accuracy has generally been quite high 
on most of the quantities studied, particularly for energetics.
In the course of this work, it has also been determined what trial function
 (starting guess, as explained in the methods section) is
necessary to obtain this accuracy.  The upper-bound property of diffusion Monte Carlo
has been critical in this success.  By this, we have also learned that
the electronic correlation in transition metal oxides is entangled with
the d-p orbital hybridization in these materials.




\section{Quantum Monte Carlo}

The most common flavors of Quantum Monte Carlo that have been used on TMO's are 
variational, diffusion, and reptation Monte Carlo (VMC, DMC, and RMC, respectively).  We will summarize them 
here; one can find a more complete review in Ref \cite{Foulkes_review}. Another flavor, auxillary field Monte Carlo\cite{zhang_afqmc},
has been used in a few calculations, but will not be discussed here.

VMC is a direct application of the variational theorem.  We write the many-body wave function as a function
of many-body coordinates $\bR=[ \bbr_1,\bbr_2,\ldots,\bbr_{N_e}]$ and a set of variational parameters $\bP$.
One then approximates the ground state wave function by minimizing the energy expectation value
\begin{equation}
E(\bP)=\int \Psi^*(\bR,\bP) H \Psi(\bR,\bP) d\bR,
\end{equation}
 assuming that the wave function is normalized.  For a complicated variational ansatz such as we will introduce later, this integral cannot be evaluated
analytically.  One can, however, evaluate it using Monte Carlo by rearranging the 
integral to read 
\begin{equation}
E(\bP)=\int |\Psi(\bR,\bP)|^2 \frac{H \Psi(\bR,\bP)}{\Psi(\bR,\bP)} d\bR.
\end{equation}
Since $|\Psi(\bR,\bP)|^2$ is a probability distribution function, one can sample it using Markov chain Monte Carlo
and evaluate the energy expectation value as an average over the local energy $E_L(\bR)=\frac{H\Psi(\bR)}{\Psi(\bR)}$.
The lowest-energy approximate wave function is then found by minimizing the energy.  In practice, a combination
of energy and the variance of the local energy\cite{umrigar_optimization2} or variance only\cite{umrigar_varopt}
is optimized.

Many wave functions can be used with VMC, since the only requirement is that one can evaluate the wave function
and its derivatives quickly.  For the work covered in this article, we start with a Slater determinant of one-particle orbitals, $D$, or a linear combination of Slater determinants.  We then multiply $D$
 by the explicitly correlated inhomogeneous Jastrow correlation factor $e^U$ to obtain 
the Slater-Jastrow variational wave function $De^U$.  We write 
\begin{equation}
U=  \sum_{ijI} u(r_{iI},r_{jI},r_{ij})
\end{equation}
where the lower case indices stand for electronic coordinates, and the upper case indices are ionic
coordinates.  There is considerable choice on how to expand $u$; for concreteness, we show
one expansion that performs well enough and has been applied to TMO's.
The correlation factor is expanded in the Schmidt-Moskowitz form\cite{schmidt:4172}:
\begin{eqnarray*}
u(r_{iI},r_{jI},r_{ij})=\sum_k c_k^{ei}a_k(r_{iI}) + \sum_m c_m^{ee} b_k(r_{ij}) \\
+ \sum_{klm} c_{klm}^{eei} (a_k(r_{iI})a_l(r_{jI})+a_k(r_{jI})a_l(r_{iI}))b_k(r_{ij}),
\end{eqnarray*}
where the $a_k$ and $b_k$ functions are written as 
\begin{equation}
 \frac{1-z(r/r_{cut})}{1+\beta_k z(r/r_{cut})},
\end{equation} with different $\beta_k$ for the different types of functions. The polynomial
$z(x)=x^2(6-8x+3x^2)$ is chosen so the functions go smoothly to zero at $r_{cut}=$7.5 bohr.
The $\beta_k$'s and all the expansion coefficients $c^{ei}$,$c^{ee}$, and $c^{eei}$ are optimized.  
If there are multiple determinants, their coefficients can also be optimized.
We then use the VMC wave function as a trial function for RMC or
DMC.  

DMC and RMC are based on the so-called imaginary time Schr\"odinger equation
\begin{equation}
-\frac{d\Psi(\bR,\tau)}{d\tau}=(H-E_0)\Psi(\bR,\tau),
\label{eqn:imagtime_diff}
\end{equation}
which has a steady-state solution $\Phi_0$, the lowest energy eigenfunction with eigenvalue $E_0$
as long as $\Psi(\bR,0)$ has a non-zero overlap with $\Phi_0$. All 
non-steady-state solutions converge exponentially to the eigenstate $\Phi_0$ as $\tau$ 
goes to infinity.  Transforming to an integral equation, we have
\begin{equation}
\Phi_0(\bR_1)=\lim_{\tau \to \infty} \int G(\bR_1,\bR_0,\tau) \Psi_T(\bR_0) d\bR_0,
\label{eqn:imagtime}
\end{equation}
where $G$ is the Green's function of the imaginary time Schr\"odinger equation and $\Psi_T(\bR_0)$
is the trial wave function that we obtain from VMC.
Solving for the exact $G$ for large $\tau$ is as difficult as solving for $\Phi_0$, so we choose
some constant small value of $\tau$ for which we know $G$ accurately (for example, see
Refs \cite{Foulkes_review,unr}), and compound the operations
(suppressing the $\tau$ dependence of $G$):
\begin{equation}
\Phi_0(\bR)=\lim_{n \to \infty} \int  G(\bR,\bR_n)\ldots G(\bR_1,\bR_0)\Psi_T(\bR_0) d\bR_0 d\bR_1 \ldots  d\bR_n.
\end{equation}
Each application of $G$ is interpreted as a stochastic process, in the same way that the 
diffusion equation can be mapped onto Brownian particles and vice versa (in fact, for a 
free particle, the Hamiltonian is $-\frac{1}{2}\nabla^2$ and the simulation is a diffusion 
process).

 DMC performs a simulation of these random particles for large $n$.  All implementations
of DMC use a particularly clever importance sampling transformation by multiplying
the imaginary time Schr\"odinger equation (Eqn \ref{eqn:imagtime_diff}) by the 
trial function $\Psi_T(\bR)$ and working with the time-dependent function
 $\Psi_T(\bR)\Psi(\bR,\tau)$.  Since the time dependence is the same, it eventually obtains 
samples distributed according to the probability distribution function
 $P_{R_\infty}(\bR)=\Phi_0(\bR)\Psi_T(\bR)$.  This transformation improves the 
efficiency of the calculation by several orders of magnitude\cite{Foulkes_review} by
using information that we already have about the ground state in the form of a trial 
function.  The final probability distribution function
can be used to evaluate the ground-state energy as follows:
\begin{equation}
\langle E_0 \rangle = \int d\bR \Psi_T(\bR)\Phi_0(\bR) \frac{H\Psi_T(\bR)}{\Psi_T(\bR)},
\end{equation}
since $\Phi_0$ is an eigenstate of $H$ and $H$ can operate forwards or backwards.
 Any operators that do not commute with the Hamiltonian will have expectation values that are 
biased, only becoming unbiased in the limit of $\Psi_T=\Phi_0$.

We can remove the error in these operators by using reptation Monte Carlo\cite{Baroni_RMC, pierleoni_rmc}, where the random walk is performed in the space of paths: 
$s=[\bR_0, \bR_1, \ldots,\bR_{n-1}, \bR_n]$.  We sample the path probability distribution
\begin{equation}
\Pi(s)=\Psi_T(\bR_0) G(\bR_0,\bR_1)\ldots G(\bR_{n-1},\bR_n) \Psi_T(\bR_n)
\end{equation}
This can be interpreted in several different ways.  If we examine the distribution at
$\bR_0$, we can view the samples of Green's functions as acting on $\Psi_T(\bR_n)$, 
and therefore $P_{R_0}(\bR_0)=\Psi_T(\bR_0)\Phi_0(\bR_0)$.  This is the same distribution as
we obtain in DMC as the path length goes to infinity.  Alternatively, 
since $G$ is symmetric on exchange of the two $\bR$ coordinates, the probability distribution 
of $\bR_n$ is the same.  Finally, we can split the path in two, one projecting on 
$\Psi_T(\bR_0)$, and the other projecting on $\Psi_T(\bR_n)$.  We then have 
\begin{eqnarray*}
P_{R_{n/2}}(\bR_{n/2})=( G(\bR_{n/2},\bR_{n/2-1})\ldots G(\bR_1,\bR_0) \Psi_T(\bR_0) )\\
\times ( G(\bR_{n/2},\bR_{n/2+1})\ldots G(\bR_{n-1},\bR_n) \Psi_T(\bR_n) ) \\
= \Phi_0^2(\bR_{n/2})
\end{eqnarray*}
for $n \rightarrow \infty$, which allows us to obtain correct expectation values of
operators that do not commute with the Hamiltonian.

\section{Geometry optimization}

In TMO materials, it is particularly useful to be able to optimize the geometry 
of the system within QMC.  The usual way of doing this in mean-field calculations is to calculate the
forces on the atoms and use one of many minimization routines.  Unfortunately, there are not yet any 
reliable methods to calculate the force within diffusion Monte Carlo, despite much 
work in that direction\cite{pierleoni_rmc,filippi_force, assaraf_force,chiesa_force}.  
These methods all require high-accuracy trial wave functions, which we usually do not have for 
transition metals. Thus,
with the current state of the art, we are only able to optimize a few key degrees of 
freedom using the total energies from DMC calculations and line minimization.
Even this must be done carefully because of the statistical uncertainty in the DMC energy.
  What follows is the scheme used in the work presented here, which has been found to be
quite robust.

According to Bayes' theorem, given a model $M$ and a set of 
data $D$, the probability of the model given the set of data is 
 \begin{equation}
P(M|D)=\frac{P(D|M)P(M)}{P(D)}.
\end{equation}
  $P(D)$ is an unimportant normalization constant and
$P(M)$ is called the prior distribution, which we are free to set to reflect the 
{\it a priori} probability distribution on the set of models. One usually 
 sets $P(M)=1$, the unbiased maximum entropy/least knowledge condition.  In the case of normally distributed data on a 
set of points $\lbrace x_1,x_2,...,x_N \rbrace $, 
\begin{equation}
P(D|M) \propto \exp[-\sum_i (M(x_i)-D(x_i))^2/2\sigma^2(x_i)], 
\end{equation}
where $\sigma(x)$ is the statistical uncertainty of $D(x)$.

For example, in the case of bond lengths, we can limit our space of models to $M(x)=c_1+c_2x+c_3x^2$, for $x$ close to 
the minimum bond length.  This is equivalent to setting the prior distribution equal to one for all quadratic
functions and to zero for non-quadratic functions.  One then calculates several data points $D(x)$ with statistical 
uncertainties $\sigma(x)$.  The probability distribution function of the bond length $b$ is then obtained by calculating the marginal distribution
\begin{equation}
p(b)=\frac{\int \delta(-c_2/2c_3-b) P(D|M)P(M) dc_1dc_2dc_3}{\int P(D|M)P(M) dc_1dc_2dc_3}.
\end{equation}  This integral is only three-dimensional, and as such could be calculated by a grid method, but it is convenient to calculate it by Monte 
Carlo, by sampling $P(D|M)P(M)$ and binning the bond length.  
The probability distribution function for the bond length is typically a Gaussian function to 
high accuracy, so it can be described as a mean value with a statistical uncertainty.

To make this scheme more efficient, we would like to calculate QMC energies as far away from 
the minimum as possible while still maintaining accuracy.  This is because the energy changes
much more quickly far from the minimum, which mitigates the stochastic uncertainties.  That 
is, the energy scale is larger far from the minimum, so less precision is necessary.  Thus, we
should use a fitting function that is valid as far from the minimum as possible, while containing
as few parameters as possible.  For minimum energy geometries, it has been found \cite{ryo_vinet,lucas_thesis} that the Vignet or modified Morse potentials are quite good for this purpose.

\section{Approximations}

\subsection{Pseudopotentials}

In QMC, we can increase the efficiency significantly by using pseudopotentials to 
replace the core electrons with an effective potential.   This has the effect of 
removing the large fluctuations near the core, which do not contribute much to the
valence electrons' correlation, which is the important for chemical
properties.  This introduces two approximations in the technique: first, the pseudopotential 
itself, and second, the small localization error\cite{lubos_psp} in diffusion Monte Carlo.

It has been found that small-core pseudopotentials are necessary for high accuracy on 
transition metals\cite{lee_mno, dolg_psp_tm}.  On the 3d metals, which are the primary
focus in this paper, this means a Ne-core pseudopotential.  The reason for this is that
the 3d electrons occupy much the same space as the semicore 3p and, to a lesser extent, the
3s electrons.  Since the 3d electrons are strongly affected by bonding, they in turn 
interact with the semicore.  This interaction will change with correlation and chemical 
environment, so we must include the semicore electrons in accurate electronic structure
calculations.  This is not unique to QMC and is generally done in density functional theory
where high accuracy is needed\cite{ferroelectric_with_small_core}.

\subsection{Finite size errors}

When performing calculations for extended systems such as crystals, it is necessary to 
introduce periodic boundary conditions.  This is an approximation on two levels.  The
first is the standard one-body level that is corrected by using reciprocal space sampling
(i.e., k-points).  The second level is inherent in a many-body correlated method, where
the periodic boundary conditions force the electron to interact unphysically with 
its periodic image.  This is similar to the finite simulation cell error 
in classical molecular dynamics simulation. This is typically corrected by either 
modifying the Coulomb interaction to remove the spurious interaction\cite{finite_size99} or 
by an correction\cite{chiesa_sk,lucas_thesis}.  Neither of these methods has clearly been
 demonstrated 
to be superior, and both methods or similar ones have been used successfully.  Even 
with these corrections, a QMC calculation of an extended system usually involves
on the order of 40 to 100 atoms, regardless of the size of the primitive cell, followed 
by extrapolation to infinite size.

\subsection{Fixed node}

The algorithms described above are exact when the wave function can be written 
as a positive function, since then $\Psi_T\Phi_0$ is a probability distribution 
function.  For fermions, it is not usually the case that $\Psi_T$ has the same 
zeros as the exact ground state, so we make the fixed-node approximation, where
the nodal surface of the exact wave function are assumed to be the same as the 
trial wave function.  This approximation typically results in recovering 
90-95\% of the correlation energy, and can be relaxed, but 
at the cost of exponential scaling of the system size\cite{Foulkes_review}.

Given that the pseudopotential localization approximation is usually quite small for 
energy differences\cite{casula_lrdmc}, we are mostly concerned with the fixed-node error.
The Jastrow factor does not change the nodes of the wave function, so in the method outlined
above, the nodes (and thus the final accuracy) are fixed to be the nodes of the Slater 
determinant of orbitals from the
mean-field method.  It is currently not feasible to vary the orbital expansion directly for 
a large system, since the number of parameters grows to the thousands for even moderately
sized systems.  However, partial optimizations
can be done, and, as we shall see, are very effective for transition metal-oxygen systems.

\section{TM-O molecules}

Simple molecular systems are excellent starting points for the study of transition metal oxides,
since they are small enough to study carefully in a reasonable amount of time, and are also
treatable by accurate but expensive quantum chemistry techniques like Coupled Cluster.  This 
provides an additional much-needed data point to compare accuracy of the various electronic 
structure methods.  

\subsection{Near-optimal one-particle orbitals}

\begin{figure}
\includegraphics[width=\columnwidth]{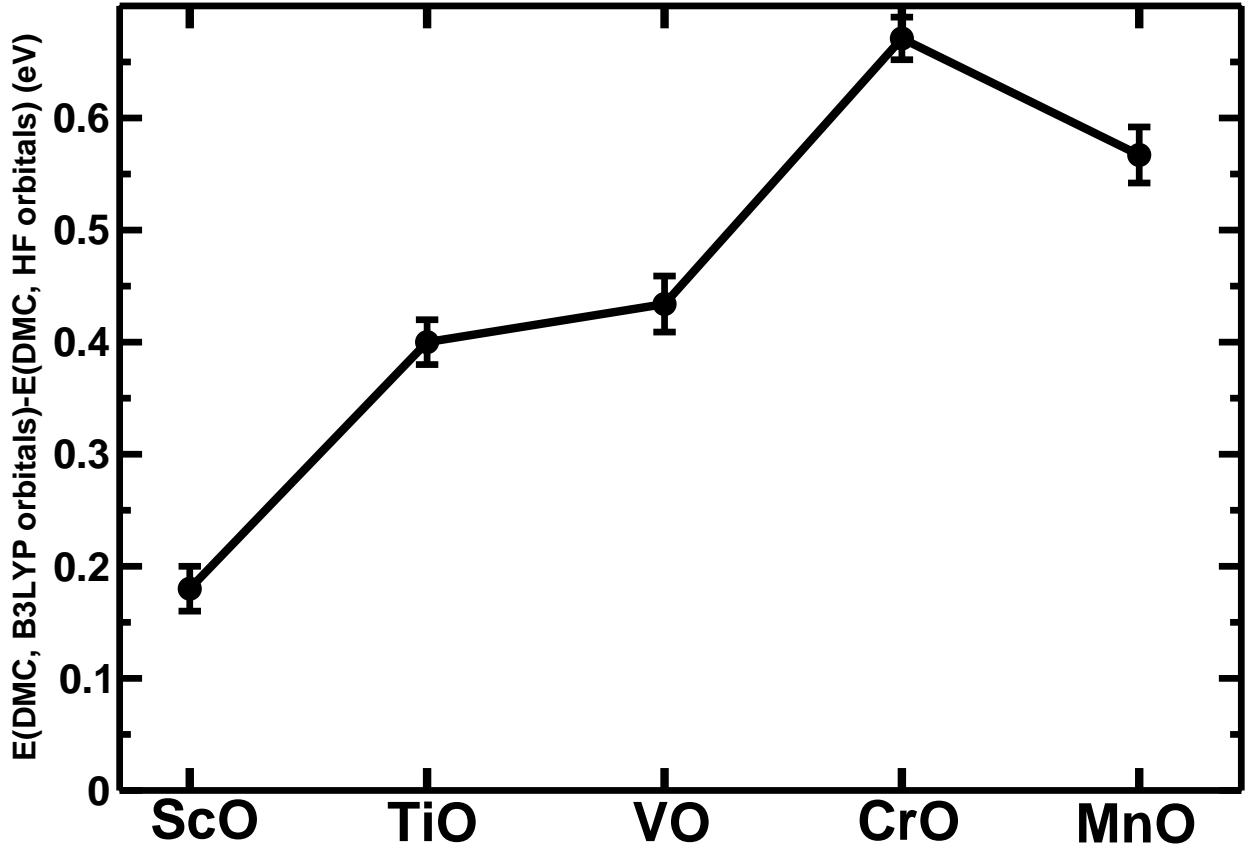}
\caption{The energy gain in DMC from using B3LYP orbitals as a function of the metal monoxide. The line is a guide to the eye. Taken from Ref \cite{wagner_jcp}.}
\label{fig:energy_gain_b3lyp}
\end{figure}

\begin{figure}
\begin{center}
\includegraphics[width=5cm]{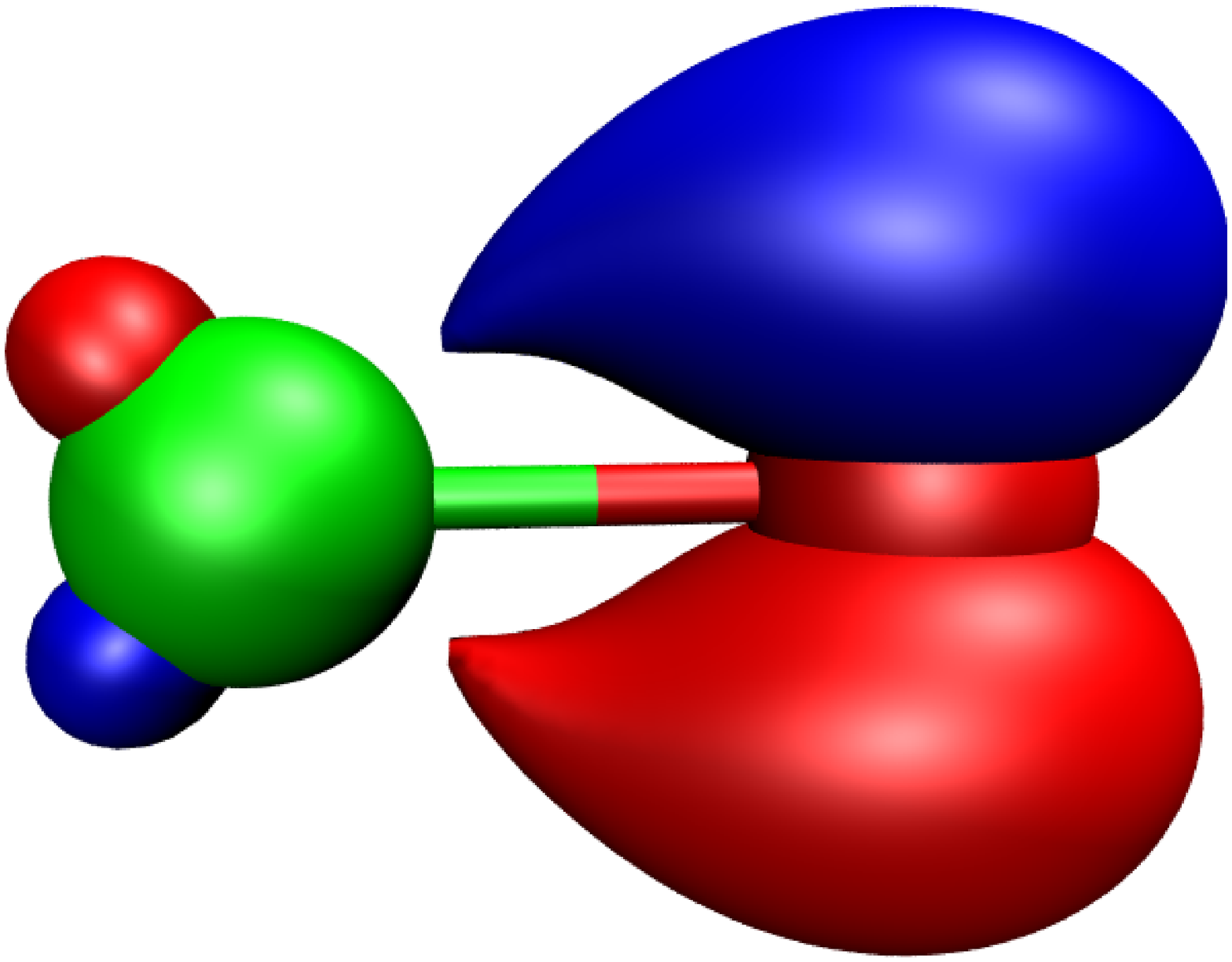} 
\includegraphics[width=5cm]{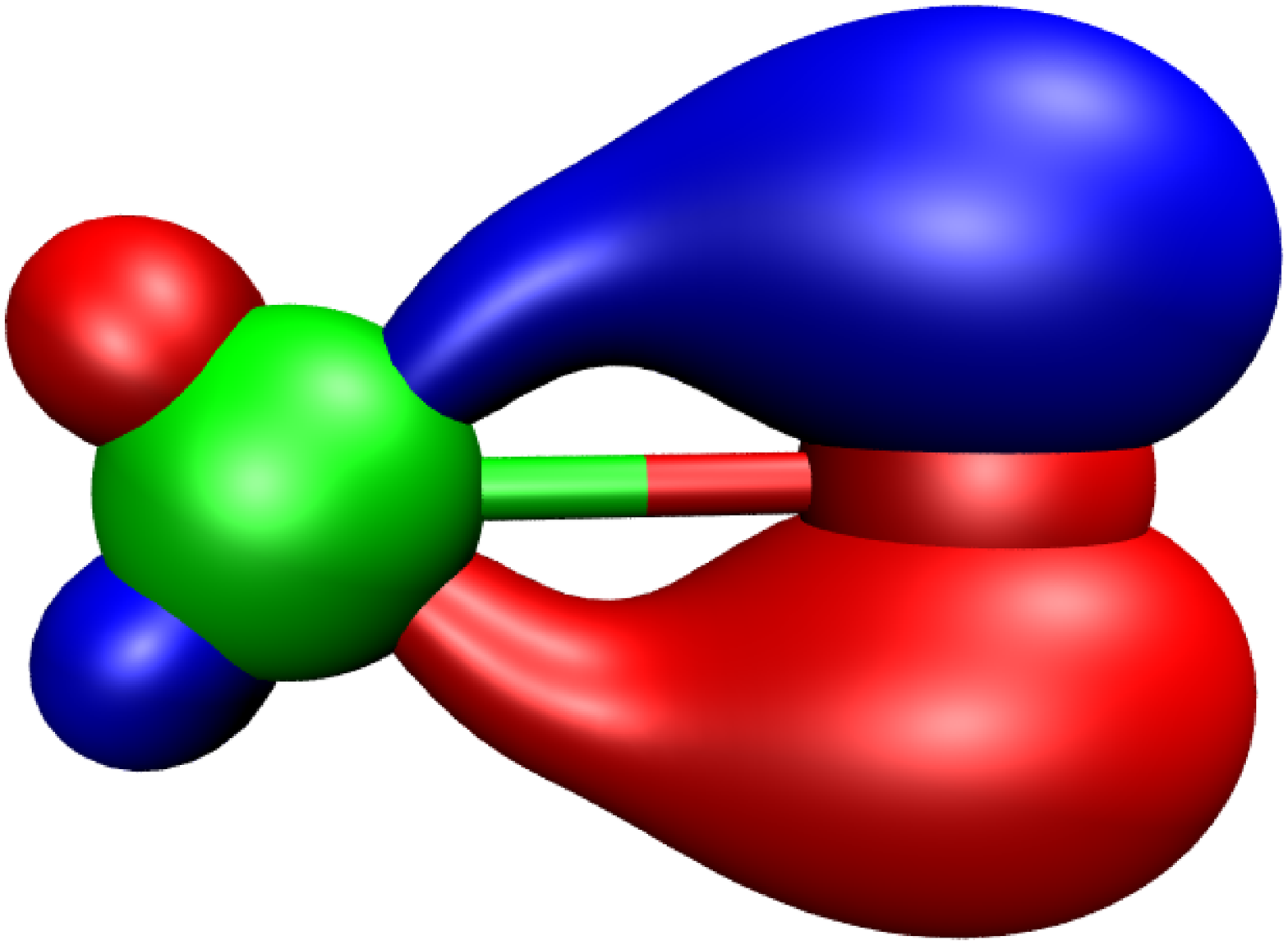}
\end{center}
\caption{The d-p hybridization orbital (doubly occupied) isosurface for TiO in Hartree-Fock (left) and B3LYP (right).
B3LYP enhances the hybridization significantly, which leads to lower energy in QMC. Figure generated
using VMD and POV-Ray\cite{vmd, povray}}  
\label{fig:tio_hf_vs_b3lyp}
\end{figure}

Wagner and Mitas\cite{CPL_lucas} performed the first calculations using DMC on 
simple two-atom transition metal oxides (TiO and MnO), and found a strong dependence of the calculated
binding energy on the orbitals used in the Slater determinant.  They used the B3LYP hybrid 
DFT/Hartree-Fock functional, and varied the percentage of Hartree-Fock mixing.  They found
the optimal percentage to be very close to the semi-empirical value fitted by Becke for his
B3PW potential\cite{becke_3parm}.  We have plotted the energy gain of B3LYP orbitals versus
Hartree-Fock for the first five transition metal monoxide molecules in 
Fig \ref{fig:energy_gain_b3lyp}.  Upon examining the orbitals, they found a large difference
in the d-p hybridization for both TiO (Fig \ref{fig:tio_hf_vs_b3lyp}) and MnO.  This is a direct
consequence of the importance of electronic correlation in transition metals.  

To understand the importance of the one-particle orbitals, one can conceptually divide the
 total energy in three parts, each described by a
different part of the wave function:
\begin{itemize}
\item One-body and antisymmetry: the Slater determinant
\item Two-body electron interaction: Jastrow factor
\item Higher orders : implicit diffusion Monte Carlo wave function.
\end{itemize}
The first part, the Slater determinant, determines the nodes of the wave function and therefore
the ultimate accuracy of the calculation.
Empirically, in materials containing only s and p-type elements, these three parts are almost 
independent of each other--the Hartree-Fock orbitals  are
close to optimal for a Slater-Jastrow wave function.  In transition metal oxides, however, 
this situation changes, and the two-body and higher interactions are strong enough to 
change the one-body part significantly.  In TMO's, this effect seems to be largely in the 
d-p hybridization between oxygen and the transition metal.

\begin{figure}
\begin{center}
\includegraphics[width=\columnwidth]{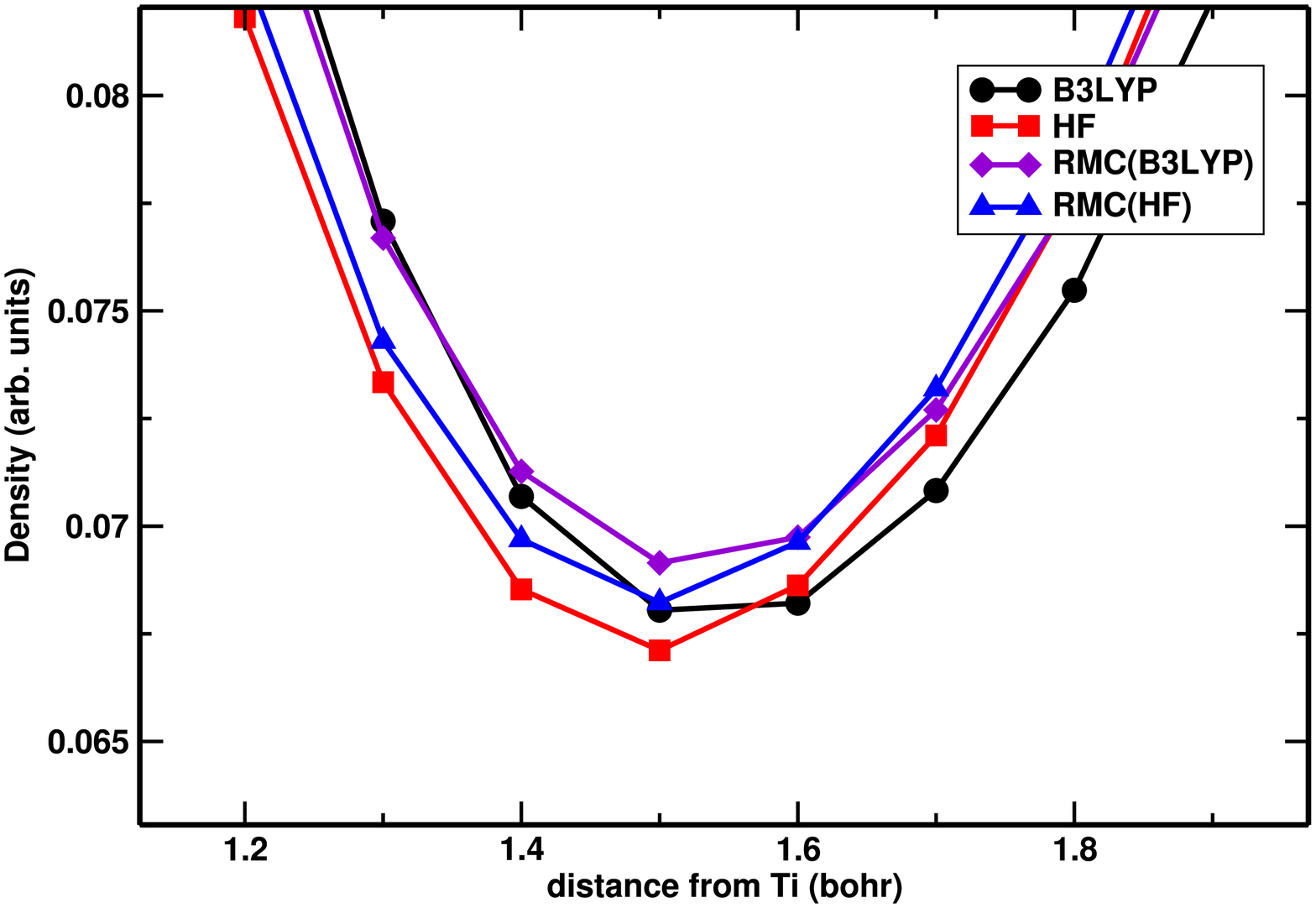}
\end{center}
\caption{The density of the Ti-O molecule projected onto the Ti-O axis in the bonding
region for various methods.  }
\label{fig:tio_dens_profile}
\end{figure}

By using the reptation Monte Carlo algorithm, we can obtain the unbiased one-particle density
within the fixed-node approximation (Fig \ref{fig:tio_dens_profile}), which gives further insight into the importance
of correlation in the one-particle density.
QMC tends to enhance the density in the bonding region (the hybridization) over both
Hartree-Fock and B3LYP, but is not able to completely repair the erroneous Hartree-Fock density
because of the fixed-node approximation.  This is the reason for the large energy gain from using
B3LYP orbitals to fix the nodal surface.

\subsection{Energetic Performance}

\begin{table}
\begin{center}
\begin{tabular}{lcccccc}
Method & ScO & TiO & VO & CrO & MnO  & RMS \\
LDA\cite{furche:044103} & 9.09 & 9.13 & 8.48 & 6.26 & 6.51 & 2.19 \\
CCSD(T)\cite{baushlicher:189} & 6.71 & 6.64 & 6.13 & 4.20 & 3.43 & 0.31 \\ 
TPSSh\cite{furche:044103} & 7.11 & 7.18 & 6.44 & 4.45 & 4.62 & 0.38 \\
DMC\cite{wagner_jcp} &  7.06(3) & 6.81(3) & 6.54(3) & 3.98(2) & 3.66(3) & 0.21\\
AFQMC\cite{al-saidi_tio} & - & 7.02(21) & - & - & 3.79(34) & - \\
Exp\cite{merer_review} & 7.01(12) & 6.92(10) & 6.44(20) & 4.41(30) & 3.83(8) & 0\\
\end{tabular}
\end{center}
\caption{Binding energies of the first five transition metal monoxides
by different theoretical methods, along with RMS deviations from the experiment(all in eV). Statistical uncertainties in units of $10^{-2}$ eV are shown in parentheses for Monte Carlo and 
experimental results. Zero point energy corrections are estimated to be much less than the uncertainty in experiment.
There are too few AFQMC data to calculate meaningful RMS values.  }
\label{table:binding}
\end{table}

The total energy of a system is quite important for determination of lowest-energy
spin states, competing phases, reactions, etc, and is a place where traditional density functional
theory has encountered difficulties on transition metal oxides.  In Table \ref{table:binding}, we
compare the binding energy obtained by DMC using B3LYP orbitals and several other methods.
We find excellent accuracy, with the RMS deviations of DMC within the experimental uncertainty for
 most materials.  CrO is the only molecule with a large deviation from experiment; however, it is not very 
far outside the experimental uncertainty. DMC is also able to consistently obtain a minimum energy bond 
length with errors below 0.01 \AA\ (Table \ref{table:bond_lengths}), better than any other 
published result.  
 
\begin{table}
\begin{tabular}{lcccccc}
Method & ScO & TiO & VO & CrO & MnO & RMS \\
LDA\cite{furche:044103} & 1.644 & 1.597 & 1.564 & 1.584 & 1.602 & 0.033 \\
CCSD(T)\cite{baushlicher:189} & 1.680 & 1.628 & 1.602 & 1.634 & 1.66 & 0.011 \\ 
TPSSh\cite{furche:044103} & 1.659 & 1.613 & 1.582 & 1.612 & 1.628 & 0.012 \\
DMC\cite{wagner_jcp} &  1.679(2) & 1.612(3) & 1.587(3) & 1.617(4) & 1.652(4) & 0.008 \\ 
Exp\cite{merer_review} & 1.668 & 1.623 & 1.591 & 1.621 & 1.648 & 0\\
\end{tabular}
\caption{Bond lengths in \AA for the first five transition metal monoxide molecules.   }
\label{table:bond_lengths}
\end{table}

\subsection{Dipole moments}

While energetics are very important for electronic structure calculations, one is also
often interested in non-energetic properties, such as dipole moments.  There has been little
work done on such things within QMC, even in the context of simpler s and p systems.
To our knowledge, the only study of dipole moments other than on TMO's is of the CO 
molecule\cite{co_dipole}.
A primary reason for this lack of calculations is that until the development of RMC, 
there has not been an easy to implement method to obtain expectation values without the
mixed-estimator bias.  The commonly used methods, pure diffusion Monte Carlo
 and forward-walking\cite{caffarel_pdmc1,caffarel_pdmc2,forward_walking} do not scale well with the 
system size\cite{assaraf_fixed_num}, since they 
suffer from increased fluctuations of weights as the number of particles increases.
One can also use extrapolated estimation, where the expectation value of an operator is estimated as
$\langle{\cal O}\rangle=2 \langle{\cal O}\rangle_{DMC} - \langle{\cal O}\rangle_{VMC}$, but that method introduces an additional 
approximation that one would like to avoid if possible.

RMC, on the other hand, scales quite well, and is easily applicable to medium-sized
systems such as TMO molecules.  As we have noticed above, the electronic correlation
and hybridization are very intertwined, and therefore, the electronic correlation and
dipole moment are also closely related.  In Table \ref{table:dipole_moments}, we report
the dipole moments for the first five transition metal monoxides using RMC with 
B3LYP orbitals.  RMC obtains dipole moments much higher than that found in experiment,
which is somewhat surprising given the high accuracy seen in energetic properties.
We will explore the fixed node approximation and its effect on the dipole moment in the
next section.

\begin{table}
\begin{tabular}{lcccccc}
Method & ScO & TiO & VO & CrO & MnO  \\
LDA\cite{furche:044103} & 3.57 & 3.23 & 3.10 & 3.41 & --  \\
CCSD(T)\cite{baushlicher:189} & 3.91 & 3.52 & 3.60 & 3.89 & 4.99 \\ 
TPSSh\cite{furche:044103} & 3.48 & 3.43 & 3.58 & 3.97 & --  \\
RMC\cite{wagner_jcp} &  4.61(5) & 4.11(5) & 4.64(5) & 4.76(4) & 5.3(1)  \\
Exp\cite{steimle_review} & 4.55 & 3.34(1)\cite{steimle_tio_03} & 3.355 & 3.88 & -- \\
\end{tabular}
\caption{Dipole moments in Debye.  The fixed-node RMC results have been obtained with a 
single determinant of B3LYP orbitals.  See text for an analysis of the errors involved
for the case of TiO.}
\label{table:dipole_moments}
\end{table}

\subsection{Beyond the Slater-Jastrow form}

In this section, we explore one of the biggest advantages of the QMC method-the ability 
to go beyond a Slater-Jastrow trial function if needed.  As we saw in the previous section, RMC with
the Slater-Jastrow trial function does not obtain dipole moments in agreement with 
experiment.  The dipole moment is very sensitive to electronic correlation, and we wish to 
perform as accurate a calculation as possible to approach the exact value.  We can 
do this in QMC by expanding the wave function in determinants.  We write the 
trial wave function as 
\begin{equation}
\Psi_T(\bR)= \left( \sum_i c_i D_i \right) e^U,
\end{equation}
where the $D_i$'s are determinants of one-particle orbitals, $e^U$ is the Jastrow factor,
and $c_i$'s are variational parameters.  These determinants and the initial
coefficients are taken from a Configuration Interaction calculation, and the coefficients
are reoptimized using Variational Monte Carlo in the presence of the Jastrow factor.

This last reoptimization step is crucial, since the DMC energy increases if the 
CI coefficients are kept constant.  This is a result of the strong correlation of 
these systems--the first order correlations are taken care of by the Jastrow factor, which
the CI calculation tries to describe (inefficiently) with determinants.  

\begin{figure}[h]
\includegraphics[width=\columnwidth]{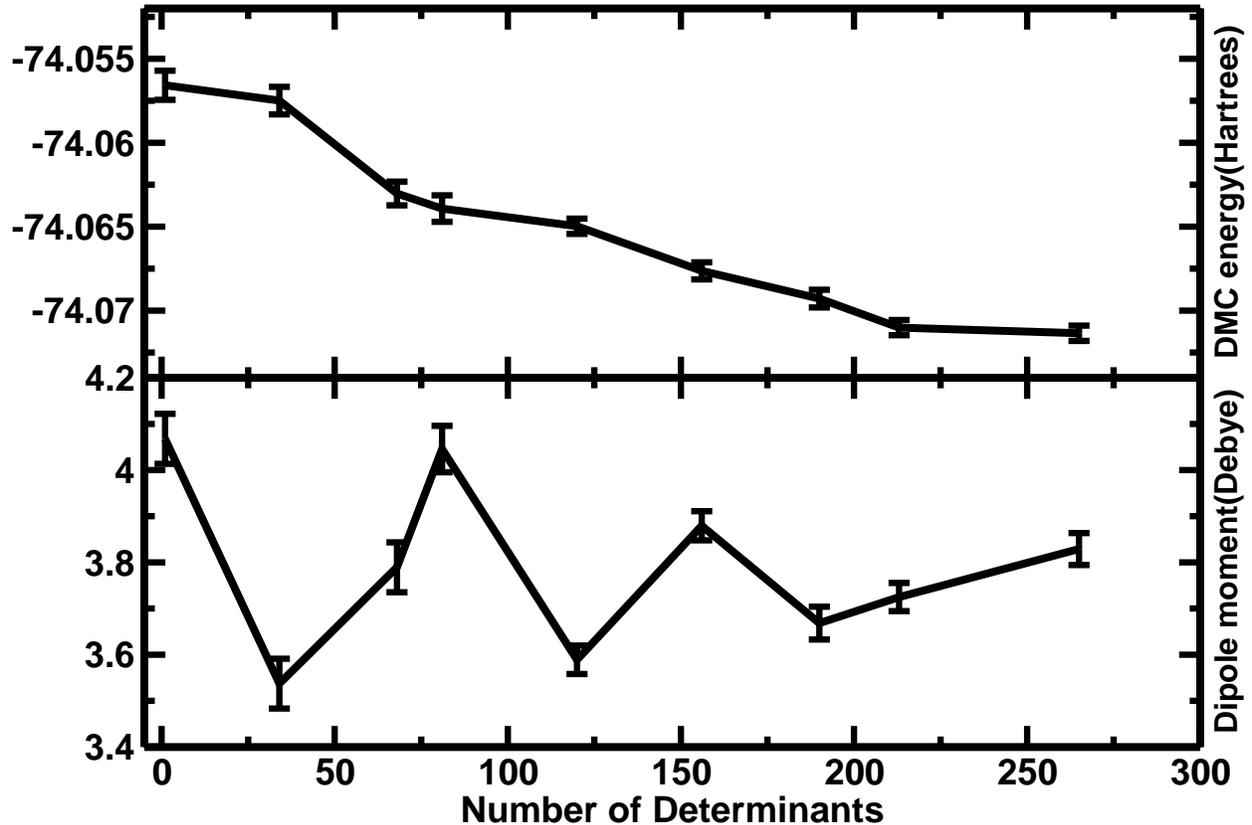}
\caption{The number of determinants versus the energy and dipole moment for TiO.The dipole moments are shifted downwards by 0.1 Debye to correct for the pseudopotential error. }
\label{fig:tio_multidet}
\end{figure}

In Fig \ref{fig:tio_multidet}, we see the convergence of this expansion for TiO.
The energy has a smooth convergence in the number of determinants, but the 
dipole moment oscillates significantly, with smaller oscillations as the
number of determinants increases.  The final result is approximately 3.8(1) Debye,
a significant change from the Slater-Jastrow trial wave function, but still quite
far from the experimental value of 3.34(1).  While this calculation is 
probably not at the exact limit, the dipole moment does not appear to change
enough to reconcile with experiment. Somewhat reassuringly, though, the 
Coupled Cluster value also predicts a larger value for the dipole moment, so 
it is possible that the experiment may be in error.  More studies of non-energy
properties using Quantum Monte Carlo are sorely needed, however, to obtain an
estimate of the expected accuracy. 

\section{Solids}

Calculations on extended TMO systems using QMC are particularly challenging, 
since QMC suffers not only from one body finite size effects (i.e., that described by 
k-point sampling), but also from many-body finite size effects, which require 
large simulation cells.  For this reason, complete studies as those
reported above for molecules are not usually attainable, and most
work is still in progress.  We will discuss a few preliminary studies and a few private
communications of work that remains unpublished at the time of this writing.  Clearly,
the details of the calculations may change, so this section is meant more as a comment
on the current state of the art.

Using QMC, there have been studies of the antiferromagnet NiO\cite{tanaka_nio,towler_nio} and 
MnO\cite{lee_mno}.  Except for Tanaka\cite{tanaka_nio}, who performed a very rough optimization 
of the lattice constant within Variational Monte Carlo, all the published studies calculated only 
the cohesive energy, which comes quite close to experiment (Table \ref{table:solid_cohesive}) 
for the materials available.  In the very recent work of 
Kolorenc and Mitas\cite{jindra_unpublished}, they obtain similar accuracy for the cohesive
energy of FeO and also obtain the correct ordering of phases for that material, which DFT
mispredicts.  In most of these materials, researchers have found a large dependence on the 
mean-field orbitals used, with the optimal orbitals ranging from Hartree-Fock to LDA.  Apparently,
there is no universal optimal mean-field method.

\begin{table}
\begin{tabular}{lccc}
Material  & DMC binding energy (eV) & Experimental & mean-field orbitals \\
NiO\cite{towler_nio} & 9.442(2) & 9.5 & Hartree-Fock\\
MnO\cite{lee_mno} & 9.40(5) & 9.5 &  \\
BaTiO$_3$\cite{lucas_unpublished}   & 31.2(3) & 31.57 & LDA \\
FeO\cite{jindra_unpublished} & 9.47(4) & 9.7 & PBE0\cite{pbe0}\\
\end{tabular}
\caption{Cohesive energies for several materials using QMC, all calculated per formula cell.
Also listed are the optimal mean-field orbitals if reported.  LDA is the local density 
approximation of DFT, and PBE0 is a hybrid functional. }
\label{table:solid_cohesive}
\end{table}

Wagner and Mitas\cite{lucas_unpublished} have also reported using the Bayesian 
optimization scheme to find the minimum energy lattice constant of BaTiO$_3$, which 
is well-known to be underestimated by over 1\% in the local density approximation 
to density functional theory, and overestimated by a similar margin in the gradient corrections.
This 1\% error in the lattice constant can affect the calculated spontaneous polarization up to 
50\%, so even this small error is not acceptable for a truely first-principles description of
this material.
DMC obtains a cubic lattice constant in error only by 0.015 $\pm$ 0.005 \AA, which is somewhat
less than half a percent, a significant improvement over the density functional results.
Also, in BaTiO$_3$, there is an energy gain in DMC of $\sim$1 eV/formula cell by using LDA orbitals
instead of Hartree-Fock orbitals, and they report that it is due to a similar change in d-p 
hybridization that is seen in the transition metal monoxide molecules.

\section{Conclusions}

On the systems that have been tested thus far, QMC offers unprecedented accuracy
in a completely first-principles and scalable method, particularly in the energetics
of the systems.
The d-p hybridization of transition metal oxides is strongly affected by 
electronic correlation.  Using QMC methods, we can clearly see this, both by 
investigating the minimum-energy orbitals and by examining the one-particle density 
and dipole moment within QMC.  The dipole moment in particular is strongly affected by 
the level of correlation present in the quantum mechanics approximation.

On TMO molecules, we have a significant gain in the total energy on expansion into determinants,
of about 0.5 eV.  This means that we are relying on cancellation of errors for the high accuracy
of QMC, although to a much lesser degree than post-Hartree-Fock approaches and DFT.  We see this
error
in the dipole moment, which does not benefit from cancellation of errors.  On the molecules,
however, we can use a brute-force approach by expanding in determinants
and come quite close to the true ground state.  However, this kind of expansion will ultimately
fail for large systems, since the number of determinants grows very quickly with system size.
In order to reliably check the QMC results, it is vital to develop new reasonably scaling 
wave functions that go beyond the Slater-Jastrow form.  Some work has been done in this direction with the RVB\cite{casula_bcs}, 
Pfaffian\cite{michal_prl}, and backflow\cite{backflow1,backflow2} wave functions in QMC. 
These wave functions' accuracy should be tested on TMO systems in the future.  Equally important
are optimization schemes within VMC that can systematically minimize the energy with respect
to the wave functions' parameters despite the stochastic nature of VMC, which is under
serious investigation\cite{umrigar_optimization2,umrigar_opt07}.  Finally, 
we need to be able to calculate forces within QMC accurately and efficiently.  The current 
state of the art is not sufficient to treat transition metal oxides\cite{lucas_thesis}, and
the Bayesian method of geometry optimization is only efficient for a few dimensions.

The future looks promising for QMC calculations of TMO solids, with the only drawback that
the calculations are very expensive on today's computers, since one must use a large supercell.
However, the scaling with system size is quite favorable, and QMC is very easy to operate in 
parallel, so it can take advantage of low-cost processors.  It has already been shown for a 
few important transition metal oxide solids that QMC can obtain binding energies and 
other energetic properties with excellent accuracy, well worth the additional cost when
high accuracy is needed.  It remains to be seen how well the method performs for non-energetic 
properties, and what sort of trial wavefunctions are necessary to obtain even higher
accuracy.

I would like to acknowledge Lubos Mitas, Jindrich Kolerenc, and Michal Bajdich for their 
support and discussions in much of the work discussed in this article, as well as E. Ertekin and V. Srinivasan
for their comments on the article itself.  I would also like
to thank a NSF Graduate Research Fellowship and NSF grant EAR-0530110 for funding during the course of this work.

\bibliographystyle{unsrt}
\bibliography{review}

\begin{thebibliography}{10}

\bibitem{qmc_ordern}
A.~J. Williamson, Randolph~Q. Hood, and J.~C. Grossman.
\newblock Linear-scaling quantum monte carlo calculations.
\newblock {\em Physical Review Letters}, 87(24):246406, 2001.

\bibitem{jeff_benchmark}
J.C. Grossman.
\newblock Benchmark quantum {M}onte {C}arlo calculations.
\newblock {\em J Chem. Phys.}, 117:1434, 2002.

\bibitem{Foulkes_review}
W.~M.~C. Foulkes, L.~Mitas, R.~J. Needs, and G.~Rajagopal.
\newblock Quantum {M}onte {C}arlo simulations of solids.
\newblock {\em Rev Mod Phys}, 73:33, 2001.

\bibitem{zhang_afqmc}
Shiwei Zhang and Henry Krakauer.
\newblock Quantum {M}onte {C}arlo method using phase-free random walks with
  slater determinants.
\newblock {\em Phys. Rev. Lett.}, 90(13):136401, Apr 2003.

\bibitem{umrigar_optimization2}
C.~J. Umrigar and C.~Filippi.
\newblock Energy and variance optimization of many-body wave functions.
\newblock {\em Phys Rev Lett}, 94:150201, 2005.

\bibitem{umrigar_varopt}
C.J. Umrigar, K.G. Wilson, and J.W. Wilkins.
\newblock Optimized trial wave functions for quantum {M}onte {C}arlo
  calculations.
\newblock {\em Phys. Rev. Lett.}, 60:1719, 1988.

\bibitem{schmidt:4172}
K.~E. Schmidt and J.~W. Moskowitz.
\newblock Correlated {M}onte {C}arlo wave functions for the atoms he through
  ne.
\newblock {\em The Journal of Chemical Physics}, 93(6):4172--4178, 1990.

\bibitem{unr}
C.J. Umrigar, M.P. Nightingale, and K.J. Runge.
\newblock A {DMC} method with small timestep errors.
\newblock {\em J. Chem. Phys.}, 99:2865, 1993.

\bibitem{Baroni_RMC}
S.~Baroni and S.~Moroni.
\newblock Reptation quantum {M}onte {C}arlo: A method for unbiased ground-state
  averages and imaginary-time correlations.
\newblock {\em Phys. Rev. Lett.}, 82:4745, 1999.

\bibitem{pierleoni_rmc}
C.~Pierleoni and D.~M. Ceperley.
\newblock {\em ChemPhysChem}, 6:1872, 2005.

\bibitem{filippi_force}
C.~Filippi and C.~J. Umrigar.
\newblock Correlated sampling in quantum {M}onte {C}arlo: A route to forces.
\newblock {\em Phys Rev B}, 61:R16291, 2000.

\bibitem{assaraf_force}
R.~Assaraf and M.~Caffarel.
\newblock Zero-variance zero-bias principle for observables in quantum {M}onte
  {C}arlo: Application to forces.
\newblock {\em J. Chem. Phys.}, 119:10536, 2003.

\bibitem{chiesa_force}
S.~Chiesa, D.M. Cepereley, and S.~Zhang.
\newblock Accurate, efficient, and simple forces computed with quantum {M}onte
  {C}arlo methods.
\newblock {\em Phys. Rev. Lett.}, 94:036404, 2005.

\bibitem{ryo_vinet}
Ryo Maezono, A.~Ma, M.~D. Towler, and R.~J. Needs.
\newblock Equation of state and raman frequency of diamond from quantum monte
  carlo simulations.
\newblock {\em Physical Review Letters}, 98(2):025701, 2007.

\bibitem{lucas_thesis}
L.~K. Wagner.
\newblock Quantum {M}onte {C}arlo for transition metal systems: Method
  developments and applications.
\newblock {\em Thesis}, 2006.

\bibitem{lubos_psp}
L.~Mitas, E.L. Shirley, and D.M. Ceperley.
\newblock Nonlocal pseudopotentials and diffusion {M}onte {C}arlo.
\newblock {\em J. Chem Phys}, 95:3467, 1991.

\bibitem{lee_mno}
J.W. Lee, L.~Mitas, and L.K. Wagner.
\newblock Quantum {M}onte {C}arlo study of {MnO} solid.
\newblock arXiv:cond--mat/0411247, 2004.

\bibitem{dolg_psp_tm}
Heinz-Jurgen Flad and Michael Dolg.
\newblock Probing the accuracy of pseudopotentials for transition metals in
  quantum {M}onte {C}arlo calculations.
\newblock {\em The Journal of Chemical Physics}, 107(19):7951--7959, 1997.

\bibitem{ferroelectric_with_small_core}
R.~D. King-Smith and David Vanderbilt.
\newblock First-principles investigation of ferroelectricity in perovskite
  compounds.
\newblock {\em Phys. Rev. B}, 49(9):5828--5844, Mar 1994.

\bibitem{finite_size99}
P.~R.~C. Kent, Randolph~Q. Hood, A.~J. Williamson, R.~J. Needs, W.~M.~C.
  Foulkes, and G.~Rajagopal.
\newblock Finite-size errors in quantum many-body simulations of extended
  systems.
\newblock {\em Phys. Rev. B}, 59(3):1917--1929, Jan 1999.

\bibitem{chiesa_sk}
Simone Chiesa, David~M. Ceperley, Richard~M. Martin, and Markus Holzmann.
\newblock Finite-size error in many-body simulations with long-range
  interactions.
\newblock {\em Physical Review Letters}, 97(7):076404, 2006.

\bibitem{casula_lrdmc}
Michele Casula, Claudia Filippi, and Sandro Sorella.
\newblock Diffusion {M}onte {C}arlo method with lattice regularization.
\newblock {\em Physical Review Letters}, 95(10):100201, 2005.

\bibitem{CPL_lucas}
L.~K. Wagner and L.~Mitas.
\newblock A quantum {M}onte {C}arlo study of electron correlation in transition
  metal oxygen molecules.
\newblock {\em Chem Phys Lett}, 370:412, 2003.

\bibitem{becke_3parm}
Axel~D. Becke.
\newblock Density-functional thermochemistry. {III.} the role of exact
  exchange.
\newblock {\em The Journal of Chemical Physics}, 98(7):5648--5652, 1993.

\bibitem{wagner_jcp}
Lucas~K. Wagner and Lubos Mitas.
\newblock Energetics and dipole moment of transition metal monoxides by quantum
  {M}onte {C}arlo.
\newblock {\em The Journal of Chemical Physics}, 126(3):034105, 2007.

\bibitem{vmd}
W.~Humphrey, A.~Dalke, and K.~Schulten.
\newblock Vmd - visual molecular dynamics.
\newblock {\em J. Molec. Graphics}, 14.1:33--38, 1996.

\bibitem{povray}
Persistence of~Vision Pt.~Ltd.
\newblock Persistence of vision (tm) raytracer.
\newblock 2004.

\bibitem{furche:044103}
Filipp Furche and John~P. Perdew.
\newblock The performance of semilocal and hybrid density functionals in 3d
  transition-metal chemistry.
\newblock {\em The Journal of Chemical Physics}, 124(4):044103, 2006.

\bibitem{baushlicher:189}
C.~W. Bauschlicher and P.~Maitre.
\newblock Theoretical study of the first transition row oxides and sulfides.
\newblock {\em Theor Chim Acta}, 90:189, 1995.

\bibitem{al-saidi_tio}
W.~A. Al-Saidi, Henry Krakauer, and Shiwei Zhang.
\newblock Auxiliary-field quantum {M}onte {C}arlo study of tio and mno
  molecules.
\newblock {\em Physical Review B (Condensed Matter and Materials Physics)},
  73(7):075103, 2006.

\bibitem{merer_review}
A.~J. Merer.
\newblock {\em Annu. Rev. Phys. Chem}, 40:407, 1989.

\bibitem{co_dipole}
F.~Schautz and H.-J. Flad.
\newblock Quantum {M}onte {C}arlo study of the dipole moment of {CO}.
\newblock {\em The Journal of Chemical Physics}, 110(24):11700--11707, 1999.

\bibitem{caffarel_pdmc1}
Michel Caffarel and Pierre Claverie.
\newblock Development of a pure diffusion quantum {M}onte {C}arlo method using
  a full generalized {Feynman--Kac} formula {I}. {F}ormalism.
\newblock {\em The Journal of Chemical Physics}, 88(2):1088--1099, 1988.

\bibitem{caffarel_pdmc2}
Michel Caffarel and Pierre Claverie.
\newblock Development of a pure diffusion quantum {M}onte {C}arlo method using
  a full generalized {Feynman--Kac} formula {II}. {A}pplications to simple
  systems.
\newblock {\em The Journal of Chemical Physics}, 88(2):1100--1109, 1988.

\bibitem{forward_walking}
{\em Monte {C}arlo Methods in Ab Initio Quantum Chemistry}.
\newblock World Scientific, Singapore, 1994.

\bibitem{assaraf_fixed_num}
Roland Assaraf, Michel Caffarel, and Anatole Khelif.
\newblock Diffusion monte carlo methods with a fixed number of walkers.
\newblock {\em Phys. Rev. E}, 61(4):4566--4575, Apr 2000.

\bibitem{steimle_review}
T.~C. Steimle.
\newblock {\em Int Reviews in Physical Chemistry}, 19:455, 2000.

\bibitem{steimle_tio_03}
T.~C. Steimle and W.~Virgo.
\newblock The permanent electric dipole moments of the {X3D, E3P, A3P and B3P}
  states of titanium monoxide, {TiO}.
\newblock {\em Chemical Physics Letters}, 381:30, 2003.

\bibitem{tanaka_nio}
S.~Tanaka.
\newblock Variational quantum monte-carlo approach to the electronic structure
  of {NiO}.
\newblock {\em J. Phys. Soc. Japan}, 64:4270, 1995.

\bibitem{towler_nio}
R.J. Needs and M.D. Towler.
\newblock The diffusion quantum {M}onte {C}arlo method: designing trial wave
  functions for {NiO}.
\newblock {\em Int. J. of Modern Physics B}, 17:5425, 2003.
\newblock also available at
  http://www.tcm.phy.cam.ac.uk/~mdt26/publications.html.

\bibitem{jindra_unpublished}
J.~Kolorenc and L.~Mitas.
\newblock {\em private communication}, 2007.

\bibitem{lucas_unpublished}
L.~K. Wagner and L.~Mitas.
\newblock {\em unpublished work}, 2006.

\bibitem{pbe0}
Carlo Adamo and Vincenzo Barone.
\newblock Toward reliable density functional methods without adjustable
  parameters: The pbe0 model.
\newblock {\em The Journal of Chemical Physics}, 110(13):6158--6170, 1999.

\bibitem{casula_bcs}
Michele Casula and Sandro Sorella.
\newblock Geminal wave functions with jastrow correlation: A first application
  to atoms.
\newblock {\em The Journal of Chemical Physics}, 119(13):6500--6511, 2003.

\bibitem{michal_prl}
M.~Bajdich, L.~Mitas, G.~Drobny, L.K. Wagner, and K.E. Schmidt.
\newblock Pfaffian pairing wave functions in electronic-structure quantum
  {M}onte {C}arlo simulations.
\newblock {\em Phys. Rev. Lett}, 96:130201, 2006.

\bibitem{backflow1}
N.~D. Drummond, P.~Lopez Rios, A.~Ma, J.~R. Trail, G.~G. Spink, M.~D. Towler,
  and R.~J. Needs.
\newblock Quantum {M}onte {C}arlo study of the ne atom and the ne+ ion.
\newblock {\em The Journal of Chemical Physics}, 124(22):224104, 2006.

\bibitem{backflow2}
P.~Lopez Rios, A.~Ma, N.~D. Drummond, M.~D. Towler, and R.~J. Needs.
\newblock Inhomogeneous backflow transformations in quantum {M}onte {C}arlo
  calculations.
\newblock {\em Physical Review E (Statistical, Nonlinear, and Soft Matter
  Physics)}, 74(6):066701, 2006.

\bibitem{umrigar_opt07}
C.~J. Umrigar, Julien Toulouse, Claudia Filippi, S.~Sorella, and R.~G. Hennig.
\newblock Alleviation of the fermion-sign problem by optimization of many-body
  wave functions.
\newblock {\em Physical Review Letters}, 98(11):110201, 2007.

\end{thebibliography}

\end{document}